\documentclass{aa}
\usepackage{xcolor}
\usepackage{array}
\newcolumntype{P}[1]{>{\centering\arraybackslash}p{#1}}
\usepackage{subfig} 
\usepackage{graphicx}
\usepackage{txfonts}
\usepackage{natbib}
\def\HI{{\ion{H}{I}}}
\def\HII{{\ion{H}{II}}}
\def\OIII{{[\ion{O}{III}]}}
\def\SII{{[\ion{S}{II}]}}
\def\NII{{[\ion{N}{II}]}}
\def\OI{{[\ion{O}{I}]}}
\newcommand{\kms}{$\,$km$\,$s$^{-1}$}

\begin{document}

\title{Embedded star formation in the extended narrow line region of Centaurus A: extreme mixing observed by MUSE}
\titlerunning{\HII\  regions in the outer filament of Centaurus A}
\author{F. Santoro \inst{1,2}\fnmsep\thanks{email: santoro@astro.rug.nl},
             J. B. R. Oonk\inst{1,3},
             R. Morganti\inst{1,2},
             T.A. Oosterloo\inst{1,2},
             and C. Tadhunter\inst{4} }

\institute{ASTRON, the Netherlands Institute for Radio Astronomy, PO 2, 7990 AA, Dwingeloo, NL.\and Kapteyn Astronomical Institute, University of Groningen, PO 800, 9700 AV, Groningen, NL.\and Leiden Observatory, Leiden University, PO Box 9513, 2300 RA Leiden, NL. \and Department of Physics and Astronomy, University of Sheffield, Sheffield S3 7RH, UK}

\date{Received 19/02/2016; accepted 12/04/2016}
 
\abstract {We present a detailed study of the complex ionization structure in a small ($\sim$250 pc) extended narrow line region (ENLR) cloud near Centaurus~A using the Multi Unit Spectroscopic Explorer. This cloud is located in the so-called outer filament of ionized gas (about 15 kpc from the nucleus) where jet-induced star formation has been suggested to occur by different studies. 
We find that, despite the small size, a mixture of ionization mechanisms is operating, resulting in  considerable complexity in the spatial ionization structure. The area includes two \HII\ regions where star formation is occurring and another location where star formation must have ceased very recently. Interestingly, the extreme Balmer decrement of one of the star forming regions (H$\alpha$/H$\beta_{\rm obs}\sim$6) indicates that it is still heavily embedded in its natal cocoon of gas and dust. At all three locations a continuum counterpart is found with spectra matching those of O/B stars local to Centaurus~A. 
The \HII\ regions are embedded in a larger gas complex which is photoionized by the radiation of the central active galactic nucleus (AGN), but the O/B stars affect the spatial ionization pattern in the ENLR cloud very locally. In particular, in the surroundings of the youngest star forming region, we can isolate a tight mixing sequence in the diagnostic diagram going from gas with ionization due to a pure stellar continuum to gas only photoionized by the AGN. 
These results emphasize the  complexity and the mixture of processes occurring in star forming regions under the influence of an AGN radiation. This is relevant for our understanding of AGN-induced star formation suggested to occur in a number of objects, including this region of Centaurus~A. They also illustrate  that these young stars influence the gas over only a limited region.
 }
 

\maketitle
%

\section{Introduction}

Active galactic nuclei (AGN) have the potential to influence the evolution of their host galaxy by injecting energy into the surrounding interstellar medium (ISM) \citep[see][for a review]{2014ARA&A..52..589H}. One of the ways in which this may occur is via the impact of radio jets and lobes. Signatures of the interaction between radio jets and the surrounding ISM have been found at both small and large scales in galaxies hosting an AGN  \citep[see][and references therein]{2012NJPh...14e5023M,2013Sci...341.1082M}.

Because of the injection of energy, AGN are often considered responsible for negative feedback (i.e. star formation quenching). However, the possibility that these effects produce positive feedback (i.e. triggering the star formation) has been suggested both by simulations  \citep[e.g.,][and references therein]{2002A&A...395L..13M,2012MNRAS.425..438G,2015arXiv151003594W} and by observations \citep{1993ApJ...414..563V,1997ApJ...490..698D,1998ApJ...502..245G,2000ApJ...536..266M,2005A&A...429..469O,2006ApJ...647.1040C}.
 
The jet-ISM interaction is a complex process that depends strongly on the conditions of the medium (e.g. its structure and density) and the characteristics of the radio plasma jet (e.g. its power, speed and collimation), as also indicated by the simulations of \cite{2011ApJ...728...29W}. Our understanding of this process is still poor and has mainly been limited by the spatial resolution of the observations and by the distance of the objects. Detailed observations can, thus, shed light on the small-scale physics of the jet-ISM interaction and help us understand this process in its entirety.

Centaurus A (Cen~A) is the nearest observed AGN~\footnote{We assume a distance of 3.8 Mpc \citep{2010PASA...27..457H} for which 1~arcmin=1~kpc} and has long been suggested as the best example of a radio jet emitted by the central AGN interacting with the ISM, and where this process may have induced star formation in the outskirts of the galaxy. The region where this interaction appears most clearly corresponds to filaments of highly ionized gas, between about 9 and 20 kpc from the nucleus, roughly aligned along the jet direction \citep{1975ApJ...198L..63B,1981ApJ...247..813G,1991MNRAS.249...91M}. These filaments are known to contain massive stars, some of them young ($\lesssim$ 4 Myr), and some of them exciting local \HII\ regions \citep{1998ApJ...502..245G,2000ApJ...538..594F,2000ApJ...536..266M,2002ApJ...564..688R, 2012MNRAS.421.1603C}. The alignment with the jet direction and the kinematics of the gas suggest that star formation may be triggered by the interaction of the AGN radio plasma with a cloud of cold gas left by a recent merger \citep{1998ApJ...502..245G,2000ApJ...536..266M,2005A&A...429..469O,2012MNRAS.421.1603C}. 


In particular, the so-called outer filament is a filament of ionized gas showing a heterogeneous morphology 15 kpc away from the galaxy nucleus. It is located within the extended narrow line regions (ENLR) of Cen~A, and close (in projection) to the jet radio plasma and a large \HI\ cloud  \citep[ see Fig.1 in][]{2015A&A...574A..89S}.
Evidence for anomalous velocities in the southern tip of the \HI\ cloud has been reported by \citet{2005A&A...429..469O}, and they interpreted it as the imprint of an interaction with the  jet.
\cite{2015A&A...574A..89S,2015A&A...575L...4S} further reinforced the idea of a direct interaction between the radio jet and the ISM by finding a match between the kinematics of the ionized and the neutral gas, suggesting that they are part of the same dynamical structure, likely shaped by the lateral and head-on interaction with the jet. 

OB stellar associations with very weak H$\alpha$ emission have been found to be distributed in a north-south direction along the eastern edge of the \HI\ cloud. In addition, a small number of luminous optically bright stars associated with knots of ionized gas are embedded in the outer filament indicating much younger \HII\ regions where star formation is still ongoing \citep[ see Fig.1 in][]{2015A&A...574A..89S}. This has been interpreted as evidence for jet-induced star formation, likely propagating from west to east across the region, possibly following the direction of penetration of the \HI\ cloud inside the area influenced by the jet \citep{2000ApJ...536..266M}.

Further study by \cite{2015A&A...574A..34S} has suggested that, in other locations the interaction with the jet is quenching the star formation instead. Coincident with the \HI\ cloud, they detected three unresolved CO clouds that, probably due to the kinetic energy of the AGN jet, are not gravitationally bound and thus do not form stars efficiently.
The overall star formation rate along the ionized gas filaments is estimated to be low and has a negligible effect on galaxy scales \citep[see][]{2000ApJ...536..266M,2005A&A...429..469O,2015A&A...574A..34S}. 

The picture of star formation induced by jet-ISM interaction has recently been questioned by \cite{2015ApJ...802...88N}. They proposed a scenario in which a broad wind would be able to ionize the gas of the \HI\ cloud and drive the star formation along the jet direction. They estimate that this wind is sustained by the starburst in the center of the host galaxy and enhanced by energy and matter driven outwards by the AGN.

We recently presented new observations of the outer filament obtained using the Multi Unit Spectroscopic Explorer  \citep[MUSE,][]{2010SPIE.7735E..08B} and discussed the morphology and kinematics of the ionized gas \citep{2015A&A...575L...4S}. Here, we use the same data to investigate the ionization of the gas in an ENLR cloud where an \HII\ region was previously found and was possibly associated with an arc-like structure reminiscent of shock-excited regions \citep{1998ApJ...502..245G,2000ApJ...536..266M}. That more than one mechanism is responsible for the ionization of the gas was already suggested by these studies. 
By using the capabilities of MUSE, our aim is  to disentangle the various structures present in this region, to characterize the ionization state of the gas in more detail, and to investigate the presence of young stellar sources and their influence on the surrounding gas. The good spatial resolution and wavelength coverage of the instrument allow us to image the ionization structure and map the contribution of the different sources of ionization in this part of the outer filament.

A detailed study of the characteristics and ionization state of the entire filament will be discussed in an upcoming paper (Santoro et al. in prep).

\begin{figure*}[]
\centering
\includegraphics[width=\hsize, keepaspectratio]{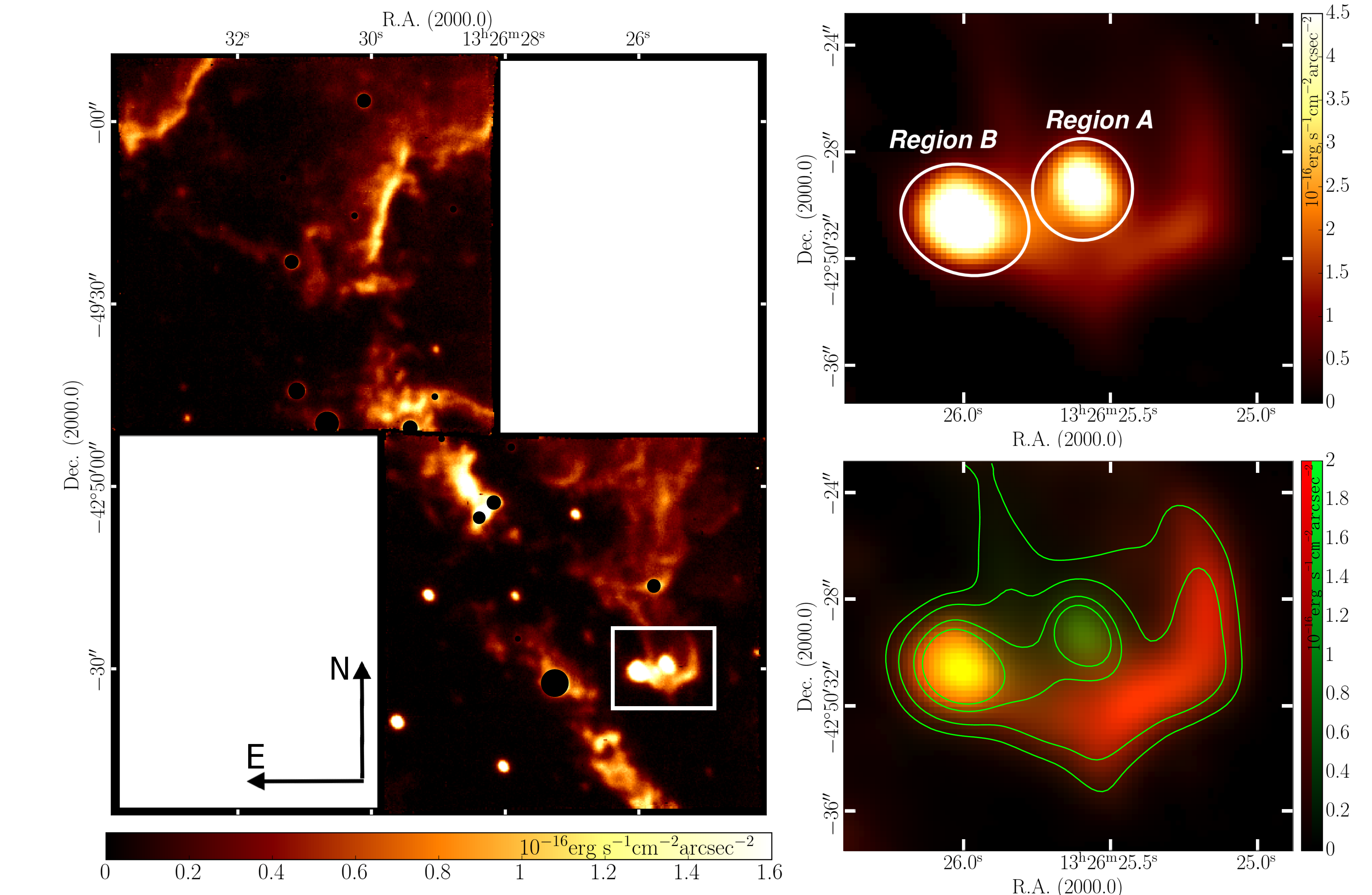}
\caption{  \textit{Left panel.} Total H$\alpha$ line flux map of the entire MUSE field of view, see Fig.1 in  \cite{2015A&A...575L...4S} for additional details. The white rectangle outline the region investigated in the present paper. \textit{Top-right panel.} Total H$\alpha$ line flux map of the region we study. Regions~A and B are outlined in white. \textit{Bottom-right panel.} Color composite image of the \OIII $\lambda$5007 (red) and H$\beta~\lambda$4861(green) emission. The color bar stretching and scale are the same for both images. H$\alpha$ intensity contours are overplotted in green. Contour levels are 2.98, 1.89, 0.75, 0.33~$\times 10 ^{-16}{\rm erg~s ^{-1}cm ^{-2}arcsec ^{-2}}$ . } 
\label{fig1}
\end{figure*}

\section{Data Reduction and Analysis}

Details about the observations we discuss here can be found in \cite{2015A&A...575L...4S}. The data have been re-calibrated using the updated ESO pipeline recipes (version 1.2.1), in combination with the command line tool EsoRex \citep{2014ASPC..485..451W}, following standard calibration steps. The only main difference is the way the sky emission was subtracted from the data. Here, we select regions across the field of view (FoV) that are free from continuum and/or line emission and create a mean sky spectrum which is then subtracted from the final cube on a pixel-by-pixel basis.

The spatial resolution of the observations is limited by the seeing, which has been estimated to be about $\sim$1 arcsec fitting the point sources across the FoV with a 2D gaussian. The data analysis presented here is carried out only in the region outlined in the left panel of Fig.\ref{fig1}. 
Before the fitting of the spectral lines is performed, the data are spatially smoothed using a Gaussian kernel with a sigma of 2.2 pixels, according to the estimated seeing level.
As shown in \cite{2015A&A...575L...4S}, the gas within this region is part of a single kinematical component and, because of this, we found no need for an extra gaussian component to model the line profiles. Only fits with a S/N $\geq$ 5 are used in order to perform the analysis described in this paper.
We use the fit of the  \OIII $\lambda$5007\AA\  line as a master fit  to model the gas distribution and kinematics: the estimated line center and width of the \OIII $\lambda$5007\AA\  line are used  to constrain the fit of the other emission lines belonging to the same spectrum.
We fit the  \OIII $\lambda$5007\AA\ , the H$\beta$ and the \NII$\lambda$5755\AA\ as single lines, the \OI$\lambda\lambda$6300,63\AA\  and the \SII$\lambda\lambda$6717,31\AA\  as doublets,  and the \NII $\lambda\lambda$6548,84\AA\ and H$\alpha$  as a triplet. In the case of a doublet/triplet we model the line emission with two/three gaussians  whose separation is fixed according to theoretical values. 

\section{The gas ionization: a unique structure}

The region of the outer filament we study here is outlined in the left panel of Fig.\ref{fig1} and has a size of about  310$\times$240  pc$^{2}$.

The H$\alpha$ line flux map in the top-right panel of Fig.~\ref{fig1} shows the presence of three structures across this region: an arc-like structure and two bright knots of emission. 
Regions~A and B in the top-right panel of Fig.\ref{fig1} correspond to the H$\alpha$ emission of the two knots, they are obtained fitting the integrated H$\alpha$ emission of each knot with a 2D Gaussian.
The color-composite \OIII$\lambda$5007 and H$\beta$ line emission image (bottom-right panel, Fig.~\ref{fig1}) shows how these morphological structures are characterized by different spectral features. The \OIII$\lambda$5007 line emission is stronger than the H$\beta$ across the arc-like structure, in line with what is found across the entire filament \citep[][Santoro et al. in prep]{1991MNRAS.249...91M,2015A&A...574A..89S}. On the other hand, the H$\beta$ emission is more prominent in region~B and dominates in region~A 
suggesting that, locally, a different/additional mechanism is playing a role in the ionization of the gas. In the following we will use line ratios to investigate in more detail and to spatially characterize the gas ionization across this region.

\subsection{The line ratios}

\begin{figure*}[t]
\centering
\subfloat[]{\includegraphics[width=0.535\textwidth]{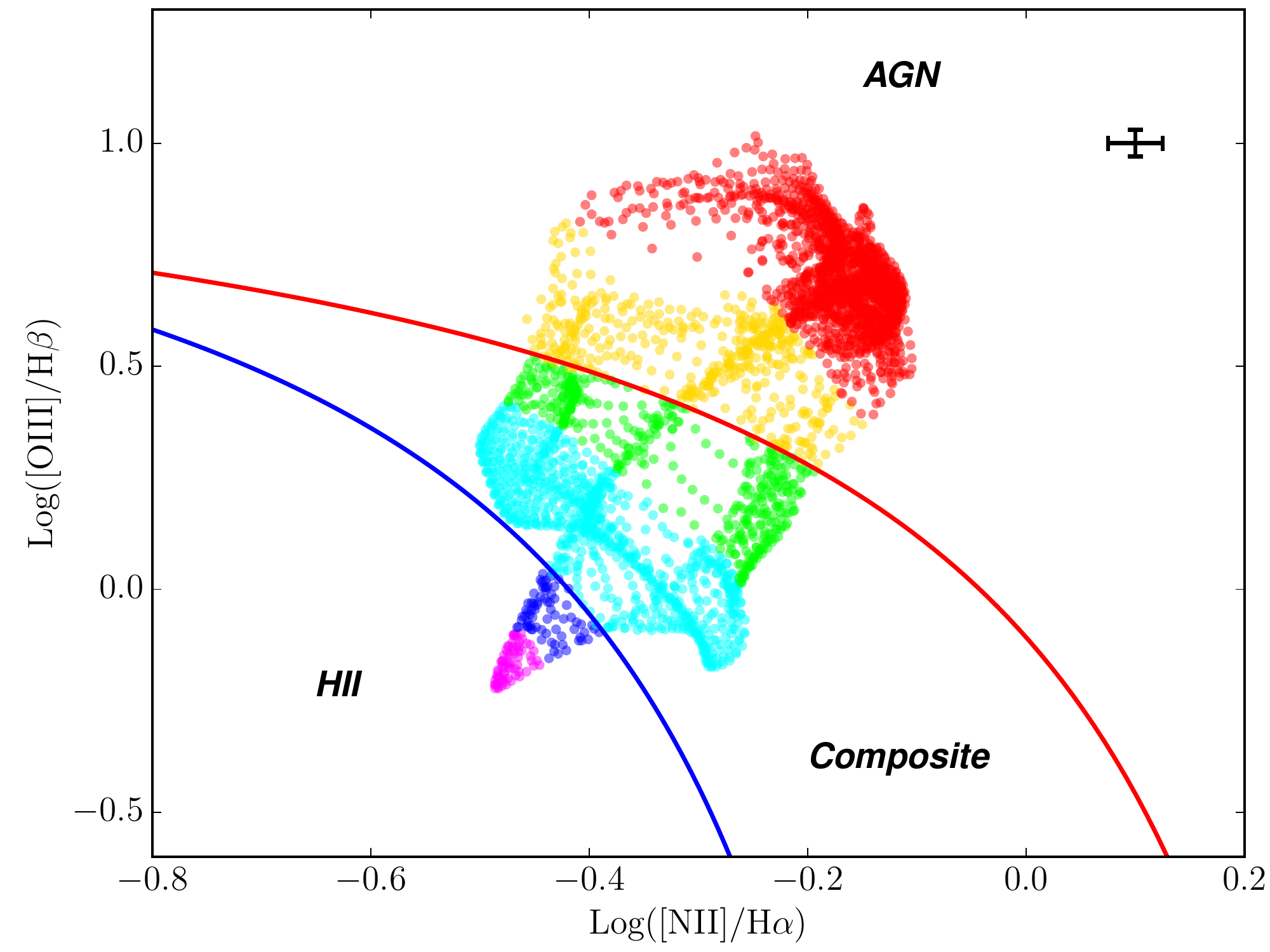}} 
\subfloat[]{\includegraphics[width=0.465\textwidth]{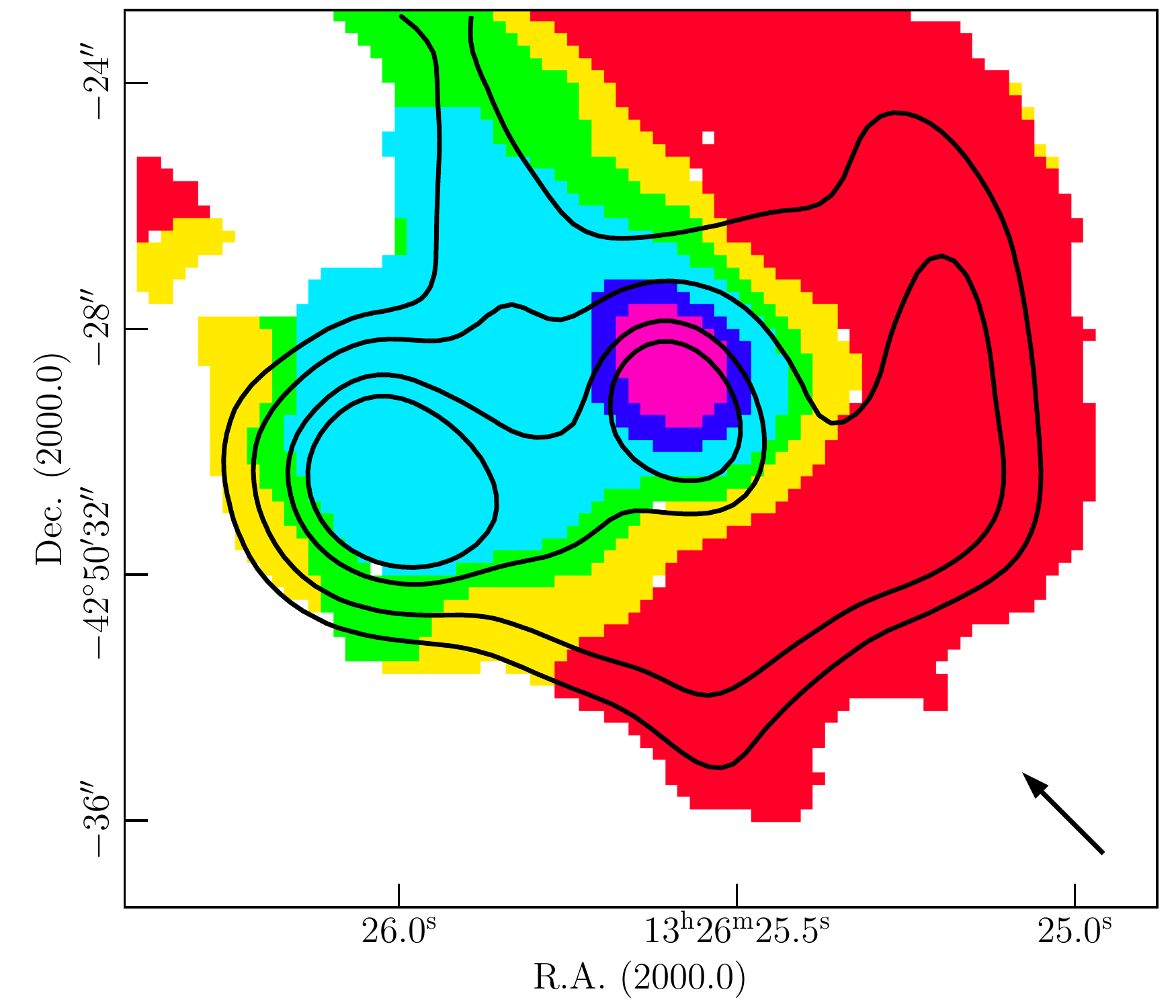}}
  %
\caption{ \textit{Left panel.} \OIII$\lambda$5007/H$\beta$ vs.  \NII$\lambda$6584 /H$\alpha$ diagnostic diagram.  The typical error on the line ratios is plotted in the upper right corner. The solid red and blue lines are the lines by \cite{2001ApJS..132...37K} and \cite{2003MNRAS.346.1055K} respectively. These lines define the \HII, the composite and the AGN region outlined in the diagram.  The points in the diagram are color coded based on the different subset selected for each of the regions. \textit{Right panel.} Pixel map of the region color coded based on the subsets defined in the diagnostic diagram. H$\alpha$ intensity contours are overplotted in black. Contour levels are 2.98, 1.89, 0.75, 0.33~$\times 10 ^{-16}{\rm erg~s ^{-1}cm ^{-2}arcsec ^{-2}}$. The arrow in bottom-right corner indicates the direction of the radio jet. }  
\label{fig2}
\end{figure*}


To discriminate between different sources of gas ionization, we use the \OIII$\lambda$5007/H$\beta$ vs.  \NII$\lambda$6584 /H$\alpha$ diagnostic diagram originally introduced by \cite{1981PASP...93....5B}.
In this diagram, the lines defined by \cite{2001ApJS..132...37K} and \cite{2003MNRAS.346.1055K} delimit three different regions where the mechanism ionizing the gas can be attributed to either radiation from young stars, to AGN continuum or to a mixture of both. We will refer to these regions as the \HII, the AGN and the composite region of the diagnostic diagram respectively.

The diagnostic diagram obtained from the fits of the spectra (left panel of Fig.\ref{fig2}) shows a significant degree of fine structure. 
Overall, the fact that the points spread over the three different regions of the diagram suggests that both the beamed ionizing continuum from the central AGN \citep[acting along the NE-SW direction, see][]{1991MNRAS.249...91M} and photoionization from newly born stars have to be taken into account to explain the gas ionization across this region of the outer filament.
It is worth mentioning that the diagnostic diagram is not significantly affected by dust, as - by design - the lines forming each ratio are close in wavelength.

To spatially characterize the gas ionization we isolate six subsets of points (two for each region of the diagnostic diagram) with increasing values of both line ratios as shown in the left panel of Fig.\ref{fig2}.  The pixel map in the right panel of Fig.\ref{fig2} clearly shows a spatially structured ionization pattern that arises from regions~A and B and spreads out. 
The ionization pattern also shows how the relative contributions of the two ionization mechanisms change over the region. The region with an \HII-like or a mixed  spectrum is small and most of the region is ionized by AGN radiation coming from the south west.

The fine structure in the ionization pattern  is particularly evident around region~A where the most extreme $\HII$-like line ratios are found. Here the ionization pattern spreads almost radially: increasing the distance from the center of region~A the photoionization by star light gradually drops and the radiation from the AGN continuum takes over the gas ionization. We thus find that the ionization of the gas has a strong dependence on the distance from the center of region~A where, based on the line ratios, we expect to find young stars. Below we model this in more detail.

\subsection{Dust and continuum sources}

\begin{figure}[t]
\centering
\includegraphics[width=\hsize, keepaspectratio]{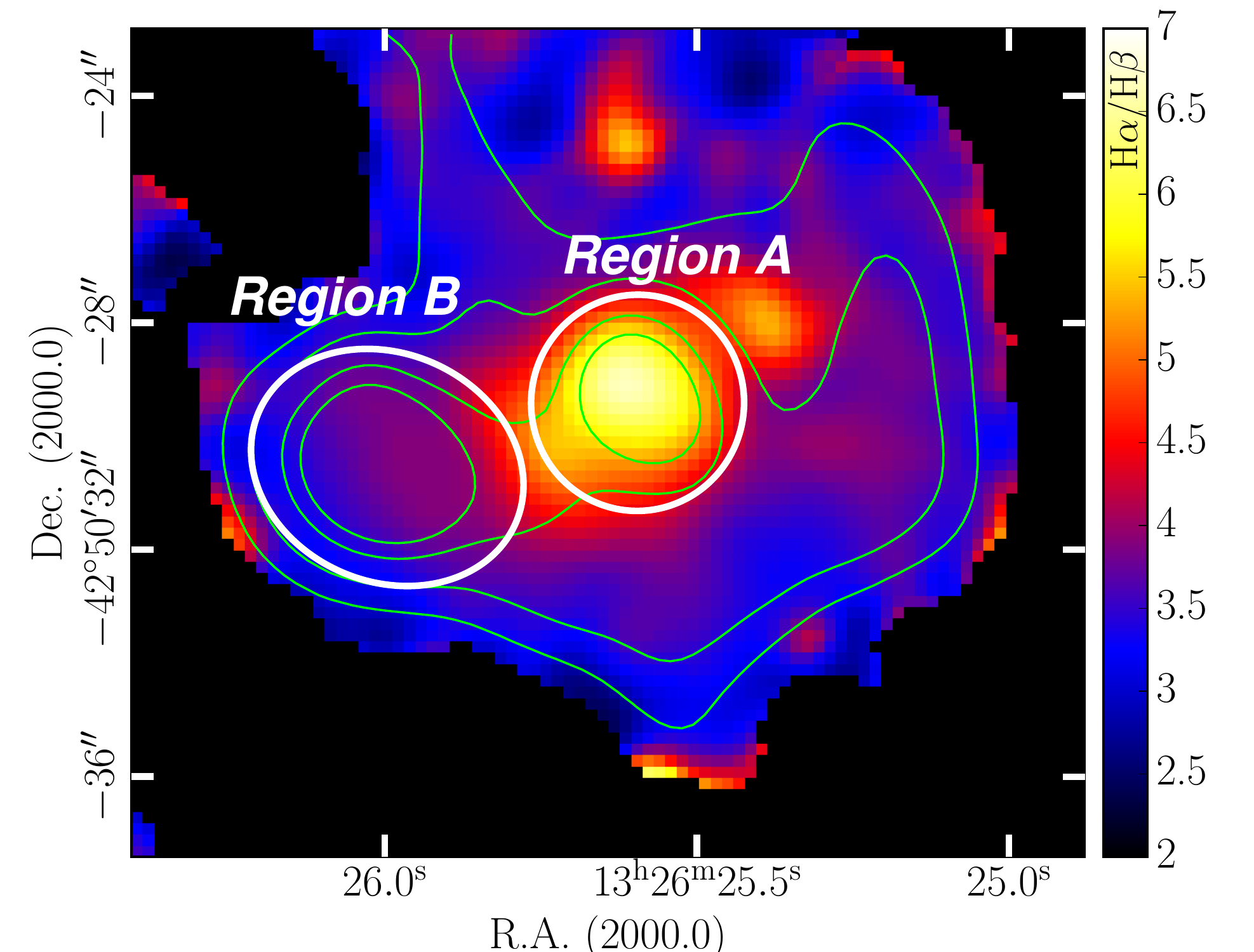}
\caption{ H$\alpha$/H$\beta_{\rm obs}$ line ratio map. Regions~A and B are outlined in white. H$\alpha$ intensity contours are overplotted in green. Contour levels are 2.98, 1.89, 0.75, 0.33~$\times 10 ^{-16}{\rm erg~s ^{-1}cm ^{-2}arcsec ^{-2}}$ .}
\label{fig3}
\end{figure}

The spectral range of MUSE allows us to trace the dust using the Balmer decrement (H$\alpha$/H$\beta$).
The line ratios show that star formation happened recently around region~B and is likely currently ongoing in region~A.
 We, thus, expect to find dust in correspondence to region~A.
In line with this, the H$\alpha$/H$\beta_{\rm obs}$ line ratio map presented in Fig.\ref{fig3} 
shows a significant increase of the Balmer decrement in region~A. 
The nebular color excess $E(B-V)$ of both regions~A and B is reported in Table \ref{Table1}. The equation we use to calculate the color excess is obtained following the approach of \cite{2013AJ....145...47M} and using the extinction curve from \cite{1989ApJ...345..245C}. 
The theoretical Balmer decrement is fixed at H$\alpha$/H$\beta$=2.86, corresponding to a temperature $T$= 10$^{4}$ K and an electron density $n_{\rm e}$ = 10$^{2}$ cm$^{-3}$ for Case B recombination \citep{2006agna.book.....O}.
The Balmer decrement in region~A is high and reaches values up to H$\alpha$/H$\beta_{\rm obs}\sim$6, this indicates that here the star formation is still embedded in its natal cocoon of gas and dust, and hence likely more recent. 

\begin{figure}[t]
\centering
\includegraphics[width=\hsize, keepaspectratio]{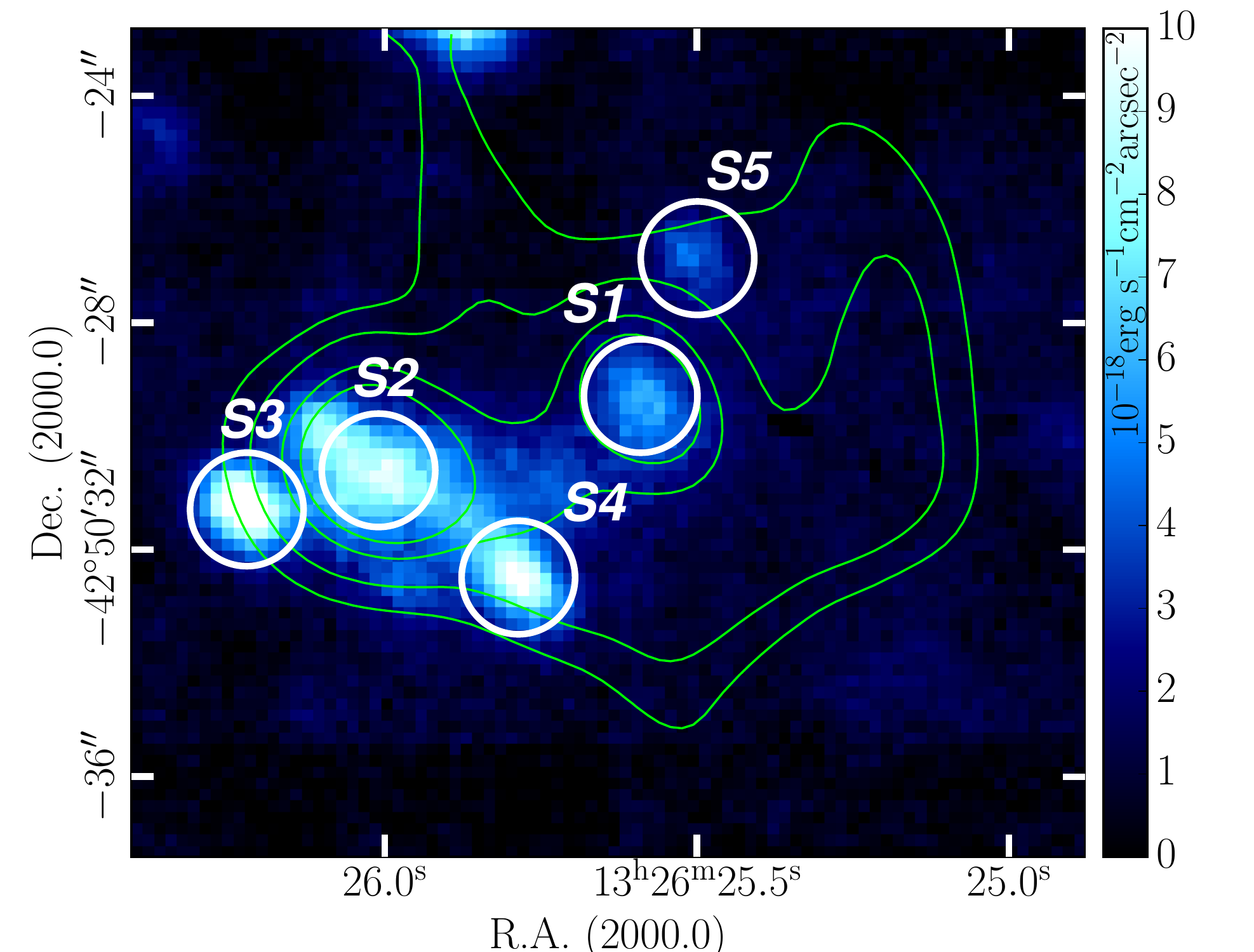}
\caption{ Continuum image obtained integrating the emission across the region in the wavelength range 4750-6650\AA\ masking line emission by the ionized gas. The position and the names of the five continuum sources are indicated in white. H$\alpha$ intensity contours are overplotted in green. Contour levels are 2.98, 1.89, 0.75, 0.33~$\times 10 ^{-16}{\rm erg~s ^{-1}cm ^{-2}arcsec ^{-2}}$.  }
\label{fig4}
\end{figure}

In the optical band, recent or ongoing star formation, if not heavily obscured by dust, should be visible through broad band continuum emission. 
Masking the gas emission lines, we integrate our datacube over the wavelength range 4750-6650~\AA\ and obtain the continuum image presented in Fig.\ref{fig4}. 
We find that the knots of ionized gas in regions A and B are associated with the continuum sources S1 and S2 (see Fig.\ref{fig4}) resembling typical \HII\ regions. Three additional continuum sources, marked as S3, S4 and S5, are found. We extract integrated spectra for the five sources and we de-redden the spectrum of the S1 source because the Balmer decrement map indicates that here the effect of the dust is relevant.
To de-redden the spectrum of S1, we use the extinction curve from \cite{1989ApJ...345..245C} and a mean Balmer decrement H$\alpha$/H$\beta_{\rm\ obs}$=6.1 extracted across the S1 region outlined in Fig.\ref{fig4}.

\begin{figure}[t]
\centering
\includegraphics[width=\hsize, keepaspectratio]{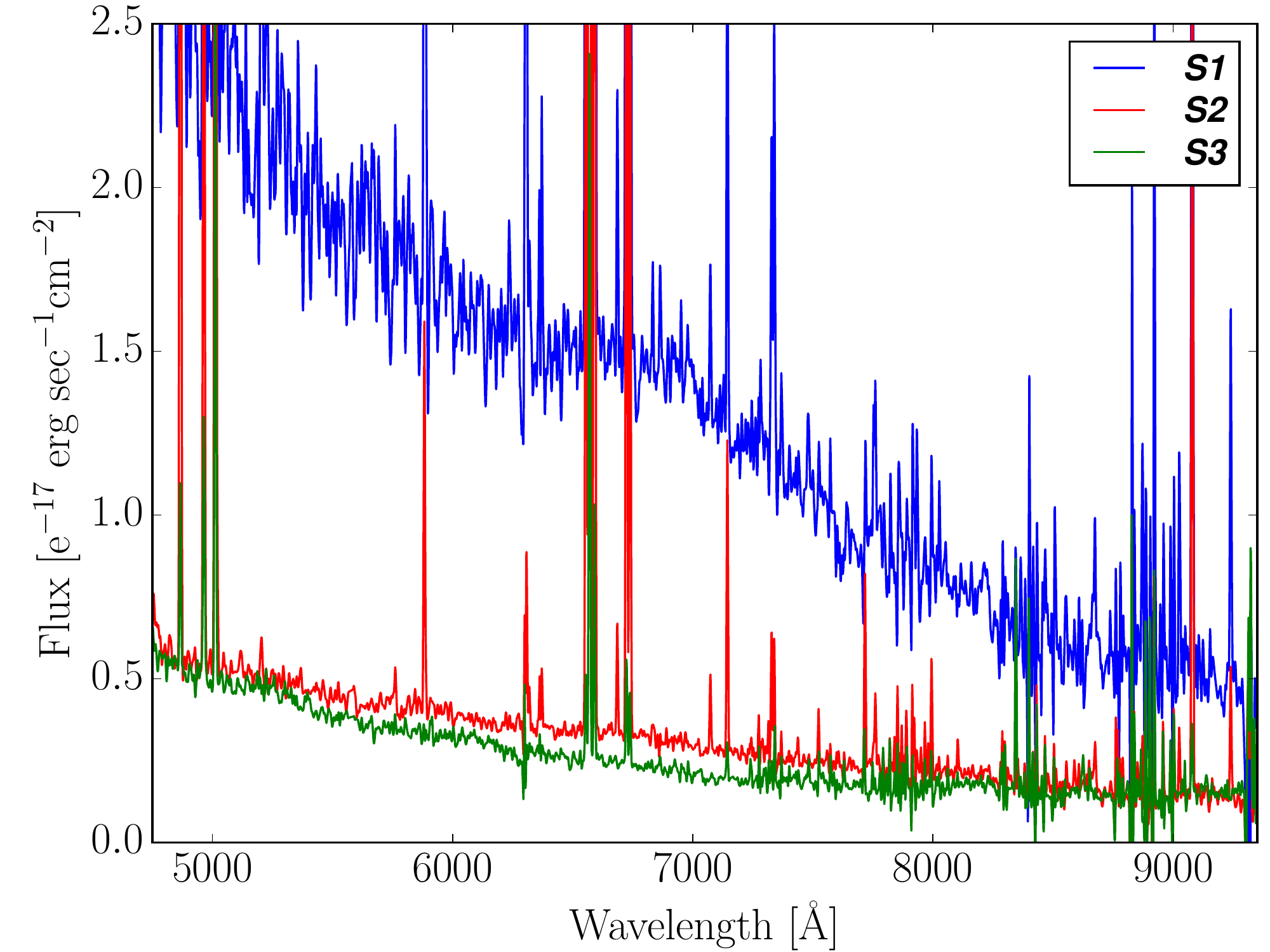}
\includegraphics[width=\hsize, keepaspectratio]{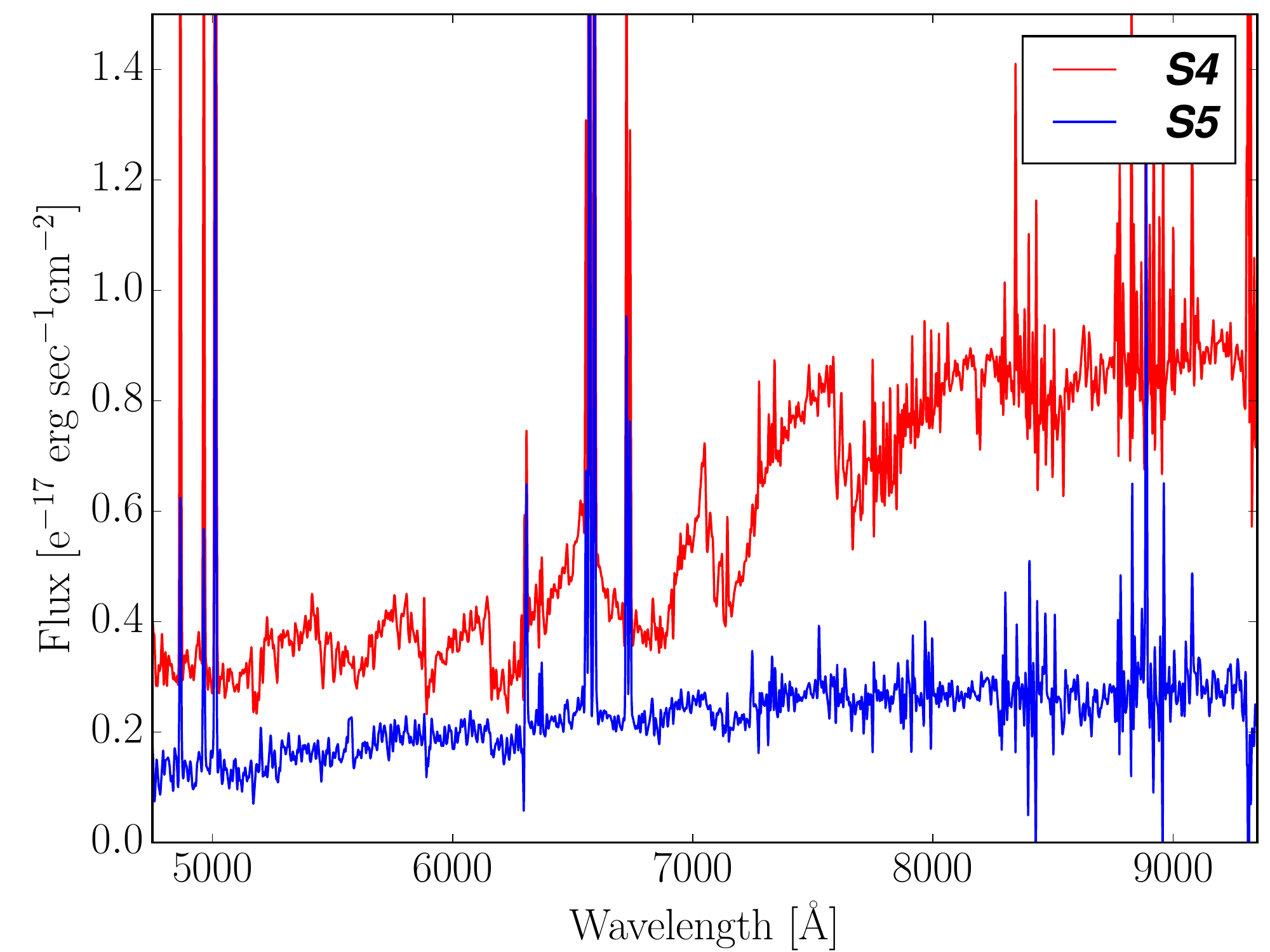}
\caption{ \textit{Upper panel.} Integrated spectra of the S1,S2 and S3 continuum sources. The spectrum of S1 has been de-reddened according to the Balmer decrement in the associated circular region. \textit{Lower panel.} Integrated spectra of the S4, S5  continuum sources. All the spectra are extracted from the circular regions associated with each continuum source and outlined in Fig.\ref{fig4}. The emission lines of the ionized gas have not been masked. Some residuals of the sky subtraction are evident in the red end of the spectra.}
\label{fig5}
\end{figure}

As shown in the upper panel of Fig.\ref{fig5}, the integrated spectra of the continuum sources S1,S2 and S3 have a blue continuum. The shape of their continuum matches with what is expected from a black body with a temperature typical of an O/B star  located at the distance of Cen~A. Contrary to what we find for S1 and S2, the continuum source S3 is not associated to a knot of ionized gas.

Two additional continuum sources with no counterpart in terms of ionized gas are found, they are marked as S4 and S5 in Fig.\ref{fig4}.
Their integrated spectra are shown in the lower panel of Fig.\ref{fig5}. 
The S4 and S5 spectra rise toward the red part of the covered spectral range and show broad absorption features, typical of M stars. Given the apparent brightness of these objects, they are likely old foreground stars, belonging to the Milky Way halo.

\subsection{The continuum sources energetics}

Using the H$\alpha$ flux of regions~A and B, we can estimate the number of ionizing photons coming from the associated continuum sources S1 and S2. 

From the diagnostic diagram presented in Fig.\ref{fig2}, we know that the AGN is affecting the gas ionization across the region. Part of the H$\alpha$ emission from regions~A and B is thus not related to starlight, but to the ionizing continuum from the AGN.
Investigating the \SII$\lambda$6717/$\lambda$6731 ratio, there is no indication of significant variations in the electron density across the region. 
We thus estimate the level of AGN contamination by extracting the mean flux of the H$\alpha$ in a region covering the arc-like structure, where we know that the ionization of the gas is mainly driven by the AGN (see the diagnostic diagram in Fig.\ref{fig2}). This value is subtracted on a pixel-by-pixel basis across both region~A and B before extracting the H$\alpha$ integrated fluxes $F(H\alpha)$ reported in Table~\ref{Table1}. 
In this way we adopt a conservative approach assuming that the contribution of the AGN to the flux of the H$\alpha$ line  is the same across the whole region.

We extract the average Balmer decrement for region~A and region~B, and estimate the related nebular color excess $E(B-V)$  and extinction $A_{H\alpha}$. The observed H$\alpha$ line flux of each region is then converted into an intrinsic luminosity $L_{\rm\ intr}$(H$\alpha$) using a distance of 3.8 Mpc.
The number of ionizing photons $Q_{0}$ coming from the continuum sources S1 and S2 is obtained using equation 5.34 given by \cite{2006agna.book.....O}. All relevant quantities and equations used to obtain the $Q_{0}$ for regions~A and B are reported in Table~\ref{Table1}.

Comparing the extracted  $Q_{0}$ values with those of synthetic stellar models \citep{1996ApJ...460..914V}, we find that the number of ionizing photons coming from the S1 and S2 continuum sources (of the order of 10$^{48-49}$ photon s$^{-1}$) is compatible with an O7 V and an O8 V star respectively.
However, massive stars usually form in associations \citep{2000prpl.conf..151C} and is thus reasonable to think that both our \HII\ regions have an underlying stellar population. 

It is worth mentioning that the line ratios sensitive to the electron density ($n_{\rm e}$) and temperature ($T_{\rm e}$) of the gas agree with the physical conditions we assumed to fix the theoretical Balmer decrement for Case B recombination \citep{2006agna.book.....O}.
In fact, the \SII$\lambda$6717/$\lambda$6731 ratio allows us to put an upper limit of $n_{e}\lesssim$100 cm$^{-3}$ while the \NII $\lambda\lambda$6548+84/$\lambda$5755 give us a $T_{e}\leq$10$^{4}$ K  \citep{2006agna.book.....O}.

\section{Photoionization models}

\begin{table*}
\centering 
\begin{tabular} { p{1.25cm} P{5cm} P{5cm} P{4.60cm} } 
\hline              
\noalign{\smallskip}
	\textbf{} 								 					&\textbf{ Region A} 			 & \textbf{ Region B} & \textbf{Equation} \\    
\hline\hline
\noalign{\smallskip} 
	\textbf{\textit{F(H$\alpha$)}}	 					& $(2.25 \pm 0.01) \times10^{-15}$ erg~s$^{-1}$~cm$^{-2}$ & $(3.79 \pm 0.01) \times10^{-15}$ erg~s$^{-1}$~cm$^{-2}$ 	&  $--$  \\  

\noalign{\smallskip} 
    \textbf{\textit{H$\alpha$/H$\beta_{\rm obs}$}} &  $5.35 \pm 0.005$ 														  &$ 3.73 \pm 0.004	$											&  $--$  \\  

\noalign{\smallskip} 
   \textbf{\textit{E(B-V)}} 								&  $0.59 \pm 0.23$ 													  & $0.25 \pm 0.09$											&  $E(B-V)=2.17 \log \dfrac{(H\alpha/H\beta)_{\rm obs}}{2.86}$  \\     

\noalign{\smallskip}          
   \textbf{\textit{A$_{\rm H\alpha}$ }} 					&  $1.48 \pm 0.58$													  & $0.62 \pm 0.16 $		  									&  $ A_{\rm H\alpha}= (2.51 \pm0.136) \times E(B-V) $ \\

\noalign{\smallskip} 
   \textbf{\textit{L$_{\rm obs}$(H$\alpha$)}} 		&  3.37$\times$10$^{36}$ erg~s$^{-1}$ 		  & 5.68$\times$10$^{36}$ erg~s$^{-1}$  & $ L_{\rm obs}(H\alpha)=F(H\alpha)4\pi d^{2} $  \\

\noalign{\smallskip} 
   \textbf{\textit{L$_{\rm intr}$(H$\alpha$)}} 			& 1.32$\times$10$^{37}$ erg~s$^{-1}$ 		  & 1$\times$10$^{37}$ erg~s$^{-1}$													& $L_{\rm int}(H\alpha)=L_{\rm obs}(H\alpha)10^{0.4A_{\rm H\alpha}} $\\

\noalign{\smallskip}    
   \textbf{\textit{Q$_{\rm 0}$}}							 & 9.6$\times$10$^{48}$ s$^{-1}$ 				  & 7.3$\times$10$^{48}$ s$^{-1}$													& $Q_{\rm 0}=7.33\times 10^{11}L_{\rm intr}(H\alpha)$  \\   
\hline 
\noalign{\smallskip}
\end{tabular}
\caption{Relevant quantities to estimate the number of ionizing photons $Q_{\rm 0}$ for regions~A and B. The last column reports the equation used in the case of extracted quantities. The distance in the calculation of the observed H$\alpha$ luminosity $L_{obs}(H\alpha)$ is assumed to be $d=3.8$ Mpc.}                                     
\label{Table1}
\end{table*}

\begin{figure}[t]
\centering
\includegraphics[width=\hsize, keepaspectratio]{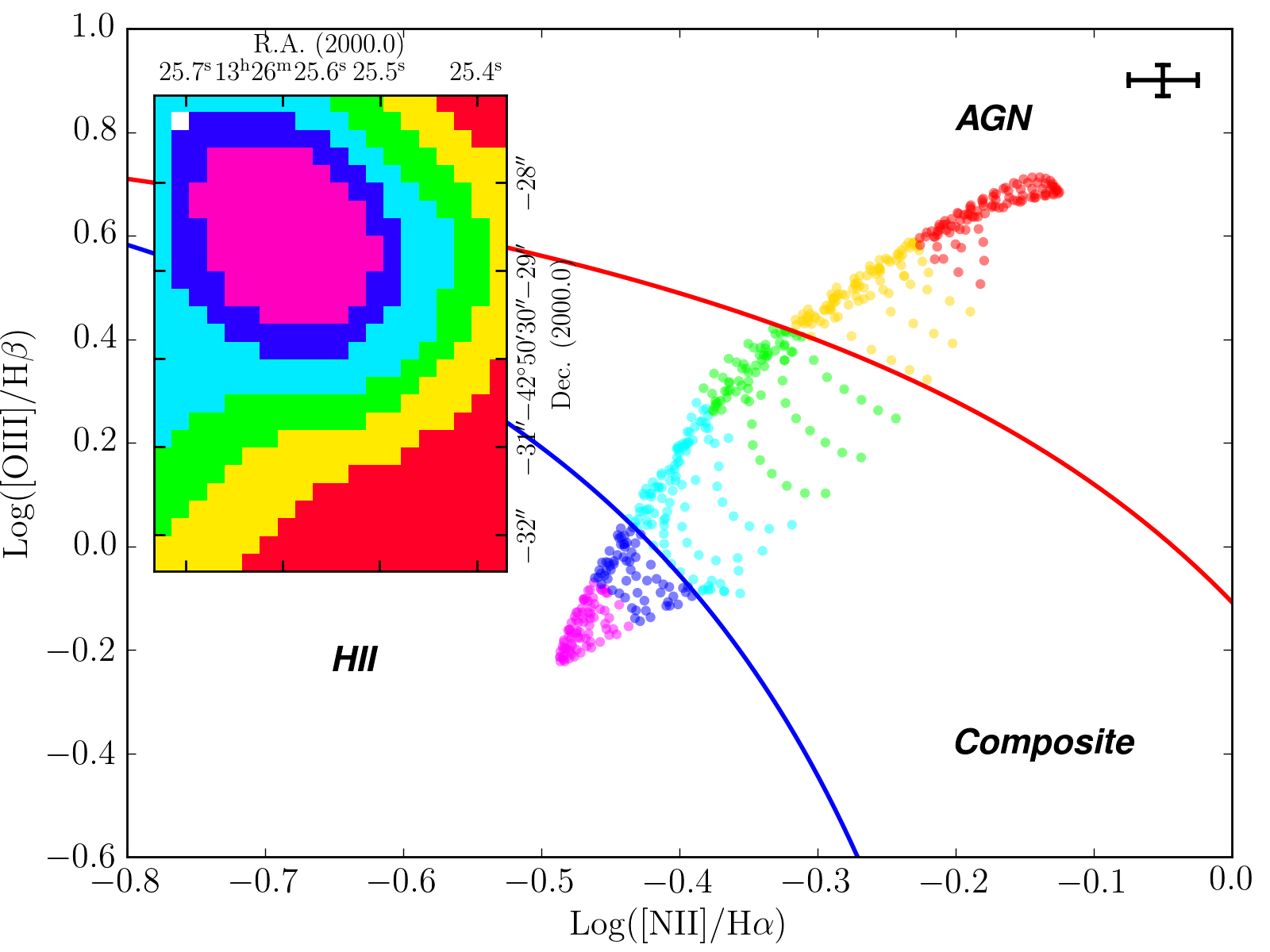}
\caption{ \OIII$\lambda$5007/H$\beta$ vs.  \NII$\lambda$6584 /H$\alpha$ diagnostic diagram showing the points related to the mixing line. The solid red and blue lines are the lines by \cite{2001ApJS..132...37K} and \cite{2003MNRAS.346.1055K} respectively. The insert in the top-left corner of the diagram is the pixel map related to the mixing sequence. The color coding of the points follows the same approach used for Fig.\ref{fig2}. } 
\label{fig6}
\end{figure} 

\begin{figure}[t]
\centering
\includegraphics[width=\hsize, keepaspectratio]{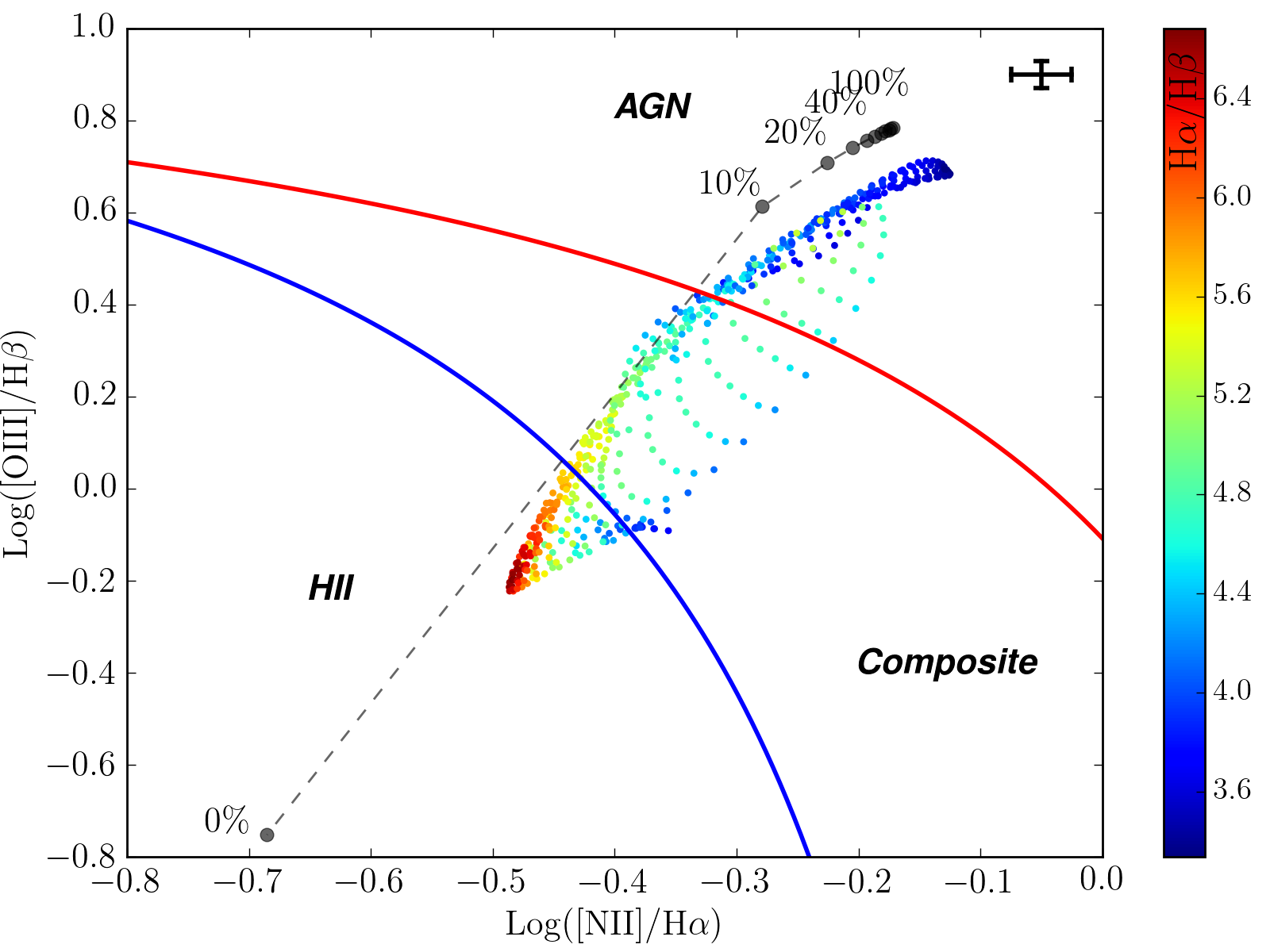}
\caption{ \OIII$\lambda$5007/H$\beta$ vs.  \NII$\lambda$6584 /H$\alpha$ diagnostic diagram showing the mixing line color coded based on the Balmer decrement. The solid red and blue lines are the lines by \cite{2001ApJS..132...37K} and \cite{2003MNRAS.346.1055K} respectively. The models are outlined with the black filled circles and connected  with the dashed black line. The point with the lower line ratios is the 100\% black body model while the one with highest line ratios is the 100\% AGN model. The fractional contribution of the AGN to the total amount of ionizing light is indicated for the models with 0-10-20-40-100\% AGN fraction. } 
\label{fig7}
\end{figure} 

From Fig.\ref{fig2} it is possible to distinguish a narrow sequence of points spreading over the three regions of the diagnostic diagram.
As shown in Fig.\ref{fig6} the points belonging this sequence, which we will call the mixing line, are all coming from the area  around the \HII\ region in region~A. In the previous section we found that here star formation is currently ongoing and still obscured by dust. The mixing line corresponds to a smooth radial transition between two ionization mechanisms and also correlates with the Balmer decrement (see Fig.\ref{fig7}).
In fact, with increasing distance form the \HII\ region both the \OIII$\lambda$5007/H$\beta$ and \NII$\lambda$6584/H$\alpha$ line ratios increase while the amount of dust decreases. 

There is a clear resemblance between the mixing line observed in our diagnostic diagram and the "extreme mixing line" predicted by the models of \cite{2001ApJS..132...37K}. 
Following their theoretical study, we use {\rm MAPPINGS III} \citep{2013ascl.soft06008S} to run ten models mixing different fractions of ionizing photons from young stars and an AGN. 
Based on our results, we tune the parameters of the models so that they can reproduce the physical conditions of the region.
Considering the $Q_{\rm\ 0}$ associated with the continuum source S1, we can model the ionizing radiation from the young stars as a single black body with temperature $T_{\rm\ BB}$=41000~K \citep[corresponding to an O7 V star, see][]{1996ApJ...460..914V}.    
Pre-computed AGN photoionization grids shows that the higher line ratios of the mixing sequence can be reproduced by an AGN
 power-law continuum with  $\alpha$=-1 and log$U$=-3, also in agreement with \cite{1991MNRAS.249...91M}.
Given the upper limit from the \SII$\lambda$6717/$\lambda$6731 line ratio, we chose an intermediate value of 40~cm$^{-3}$ for the gas electron density.
The metallicity is assumed to be solar, close to the value found by \cite{2015A&A...574A..34S} across the entire outer filament ($\sim$0.8~$Z_{\odot}$).

Starting from a model with 100\% black body contribution, we gradually increase the contribution of the AGN continuum, and scale the one of the black body, until reaching a model with 100\% AGN contribution, see Fig.\ref{fig7}. Even though there is a small offset, our simple models can qualitatively reproduce the trend of the observed mixing sequence. The mixing of photoionization by stars local to the outer filament and by the central AGN is thus a reasonable model for the ionization for this region.

While the measured line ratios are consistent with AGN photoionization for the outer part of the cloud facing the nucleus, we cannot entirely rule out a contribution from slow shocks. As discussed in \cite{2015A&A...574A..89S,2015A&A...575L...4S} the width of the emission lines across the outer filament is $\sim$80-100\kms and so models implying strong shocks are unlikely.

\section{Discussion and Conclusions}

The capabilities offered by MUSE allowed us to expand our view of the ionization structure of an ENRL cloud located in the outer filament of Cen~A and assess the role of local ionization sources.
By using the classical diagnostic diagram we have been able, for the first time, to determine the spatial structure associated with the mixing of radiation by newly born stars and the AGN continuum across this region.  

The H$\alpha$ flux map of the region we are investigating (top-right panel in Fig.\ref{fig1}) shows the presence of two knots of emission (regions~A and B) and an arc-like structure. \cite{1998ApJ...502..245G} identified an \HII\ region that corresponds to region~B and, considering the arc structure reminiscent of a shock-like feature, \cite{2000ApJ...536..266M} suggested that at least two different mechanisms (i.e. massive stars and shocks) are ionizing the gas. 
As discussed in  \cite{2015A&A...575L...4S}, and mainly because of the narrow width of the emission lines ($\sim$80-100\kms), a relatively mild and soft interaction must be taking place and the ionization of the gas is unlikely to be driven by shocks.
The study of the continuum sources and the estimate of the number of ionizing photons (see Sec.3)
confirms the presence of an \HII\ region in region~B \citep{1998ApJ...502..245G}. In addition, we identify a new \HII\ region in region~A. The continuum source S3  (see Fig.\ref{fig4}) is also a young star local to Cen~A but has no ionized gas counterpart. 

In region~A we find a significant increase in the amount of dust  (H$\alpha$/H$\beta_{\rm obs}\sim$6,  see Fig.\ref{fig3}). 
This is also supported by the far-UV observations of this region available in the GALEX archive \footnote{http://galex.stsci.edu/GR6/?page=mastform}, even though the spatial resolution of GALEX is much lower then our MUSE data. Given the presence of two continuum sources with a comparable number of ionizing photons, the fact that region~A appears less UV-bright than region~B confirms that it is more extincted. 

In our scenario, region~A represents an early evolutionary stage of star formation in this particular ENLR cloud of the outer filament, where new stars are still heavily embedded in the dusty natal cocoon. On the other hand, region~B is hosting a less young/embedded and UV-brighter \HII\ region. The young stellar source S3 found in the vicinity is not associated with a knot of ionized gas, and is likely a former \HII\ region that has already dispersed/ionized its birth-gas.

 In line with more recent star formation, the line ratios of region~A clearly show that, close to the continuum source, stellar radiation is dominating the gas ionization.
 As would be expected within the ENLR of Cen~A, the radiation of local stars is mixed with the AGN radiation field. Thanks to the  high level of fine structure in the diagnostic diagram we can map the structure of this mixing in detail.
 As shown in Fig.\ref{fig2} we find that he continuum sources in regions~A and B are driving a spatial ionization pattern that spans the entire region. 
Around region~A  the pattern is almost radial and in its inner parts we can clearly see the influence of the S1 continuum source. 
As shown in Fig.\ref{fig6}, region~A is associated with a narrow mixing line in the diagnostic diagram that also correlates with the observed Balmer decrement. 
We tried to reproduce the observed mixing line using the plasma modelling code {\rm MAPPINGS III} and mixing stellar and AGN photoionization.
Even though there is a small offset,  our synthetic mixing line is able to reproduce the trend of the observed mixing line (Fig.\ref{fig7}). This confirms that photoionization by young stars and the central AGN are likely the sources driving the gas ionization across this region.

Our models show that even a small contribution of the AGN (e.g. 10\%) can move the points of the diagnostic diagram into the AGN region (see Fig.\ref{fig7}). Therefore, it is likely that in region~A the dust shields the gas from the external radiation field of the AGN, favoring a smooth transition between stellar and AGN photoionization and increasing the chance to detect the mixing line. In this context, the fact that the AGN radiation field is acting toward the north east can also explain the offset we observe between the center of the ionization pattern and the center of the H$\alpha$ emission associated to region~A  (right panel of Fig.\ref{fig2}). 


Due to projection effects we cannot discriminate if the \HII\ regions and the arc-like structure are co-spatial or if we are observing their spatial superimposition.
However, if they were completely unrelated structures, we would not expect to observe such a smooth transition in the points of the diagnostic diagram and, more generally, a well defined ionization pattern across such a small region.

In conclusion, our results clearly show that individual clouds within ENLR can have a composite ionization and that star formation can still take place inside them coexistent with the transit of a jet and/or the radiation field from the AGN.

The effect of young stars on the ionization of the surrounding gas is very local and hence not important for the global ionization of the outer filament of Cen~A. However, we also show evidence that the conversion of gas into stars is happening, even for highly ionized gas filaments in the zone of influence of the jet/AGN ionization cone. Depending on the overall efficiency of the process this may be an important constituent for models of both positive and negative AGN feedback.

\begin{acknowledgements}
The research leading to these results has received funding from the European Research Council under the European Union's Seventh Framework Programme (FP/2007-2013) / ERC Advanced Grant RADIOLIFE-320745. Based on observations collected at the European Organisation for Astronomical Research in the Southern Hemisphere under ESO programme 60.A-9341A.  
\end{acknowledgements}

\bibliographystyle{aa}
\bibliography{DRAFT.bib}

\begin{thebibliography}{37}
\expandafter\ifx\csname natexlab\endcsname\relax\def\natexlab#1{#1}\fi

\bibitem[{{Bacon} {et~al.}(2010){Bacon}, {Accardo}, {Adjali}, {Anwand},
  {Bauer}, {Biswas}, {Blaizot}, {Boudon}, {Brau-Nogue}, {Brinchmann},
  {Caillier}, {Capoani}, {Carollo}, {Contini}, {Couderc}, {Daguis{\'e}},
  {Deiries}, {Delabre}, {Dreizler}, {Dubois}, {Dupieux}, {Dupuy}, {Emsellem},
  {Fechner}, {Fleischmann}, {Fran{\c c}ois}, {Gallou}, {Gharsa}, {Glindemann},
  {Gojak}, {Guiderdoni}, {Hansali}, {Hahn}, {Jarno}, {Kelz}, {Koehler},
  {Kosmalski}, {Laurent}, {Le Floch}, {Lilly}, {Lizon}, {Loupias}, {Manescau},
  {Monstein}, {Nicklas}, {Olaya}, {Pares}, {Pasquini}, {P{\'e}contal-Rousset},
  {Pell{\'o}}, {Petit}, {Popow}, {Reiss}, {Remillieux}, {Renault}, {Roth},
  {Rupprecht}, {Serre}, {Schaye}, {Soucail}, {Steinmetz}, {Streicher}, {Stuik},
  {Valentin}, {Vernet}, {Weilbacher}, {Wisotzki}, \&
  {Yerle}}]{2010SPIE.7735E..08B}
{Bacon}, R., {Accardo}, M., {Adjali}, L., {et~al.} 2010, in Society of
  Photo-Optical Instrumentation Engineers (SPIE) Conference Series, Vol. 7735,
  8

\bibitem[{{Baldwin} {et~al.}(1981){Baldwin}, {Phillips}, \&
  {Terlevich}}]{1981PASP...93....5B}
{Baldwin}, J.~A., {Phillips}, M.~M., \& {Terlevich}, R. 1981, \pasp, 93, 5

\bibitem[{{Bicknell}(1991)}]{1991PASAu...9...93B}
{Bicknell}, G. 1991, Proceedings of the Astronomical Society of Australia, 9,
  93

\bibitem[{{Blanco} {et~al.}(1975){Blanco}, {Graham}, {Lasker}, \&
  {Osmer}}]{1975ApJ...198L..63B}
{Blanco}, V., {Graham}, J.~A., {Lasker}, B.~M., \& {Osmer}, P.~S. 1975, \apjl,
  198, L63

\bibitem[{{Calzetti} {et~al.}(1994){Calzetti}, {Kinney}, \&
  {Storchi-Bergmann}}]{1994ApJ...429..582C}
{Calzetti}, D., {Kinney}, A.~L., \& {Storchi-Bergmann}, T. 1994, \apj, 429, 582

\bibitem[{{Cardelli} {et~al.}(1989){Cardelli}, {Clayton}, \&
  {Mathis}}]{1989ApJ...345..245C}
{Cardelli}, J.~A., {Clayton}, G.~C., \& {Mathis}, J.~S. 1989, \apj, 345, 245

\bibitem[{{Clarke} {et~al.}(2000){Clarke}, {Bonnell}, \&
  {Hillenbrand}}]{2000prpl.conf..151C}
{Clarke}, C.~J., {Bonnell}, I.~A., \& {Hillenbrand}, L.~A. 2000, Protostars and
  Planets IV, 151

\bibitem[{{Crockett} {et~al.}(2012){Crockett}, {Shabala}, {Kaviraj},
  {Antonuccio-Delogu}, {Silk}, {Mutchler}, {O'Connell}, {Rejkuba}, {Whitmore},
  \& {Windhorst}}]{2012MNRAS.421.1603C}
{Crockett}, R.~M., {Shabala}, S.~S., {Kaviraj}, S., {et~al.} 2012, \mnras, 421,
  1603

\bibitem[{{Croft} {et~al.}(2006){Croft}, {van Breugel}, {de Vries}, {Dopita},
  {Martin}, {Morganti}, {Neff}, {Oosterloo}, {Schiminovich}, {Stanford}, \&
  {van Gorkom}}]{2006ApJ...647.1040C}
{Croft}, S., {van Breugel}, W., {de Vries}, W., {et~al.} 2006, \apj, 647, 1040

\bibitem[{{Dey} {et~al.}(1997){Dey}, {van Breugel}, {Vacca}, \&
  {Antonucci}}]{1997ApJ...490..698D}
{Dey}, A., {van Breugel}, W., {Vacca}, W.~D., \& {Antonucci}, R. 1997, \apj,
  490, 698

\bibitem[{{Fassett} \& {Graham}(2000)}]{2000ApJ...538..594F}
{Fassett}, C.~I. \& {Graham}, J.~A. 2000, \apj, 538, 594

\bibitem[{{Gaibler} {et~al.}(2012){Gaibler}, {Khochfar}, {Krause}, \&
  {Silk}}]{2012MNRAS.425..438G}
{Gaibler}, V., {Khochfar}, S., {Krause}, M., \& {Silk}, J. 2012, \mnras, 425,
  438

\bibitem[{{Graham}(1998)}]{1998ApJ...502..245G}
{Graham}, J.~A. 1998, \apj, 502, 245

\bibitem[{{Graham} \& {Price}(1981)}]{1981ApJ...247..813G}
{Graham}, J.~A. \& {Price}, R.~M. 1981, \apj, 247, 813

\bibitem[{{Harris} {et~al.}(2010){Harris}, {Rejkuba}, \&
  {Harris}}]{2010PASA...27..457H}
{Harris}, G.~L.~H., {Rejkuba}, M., \& {Harris}, W.~E. 2010, \pasa, 27, 457

\bibitem[{{Heckman} \& {Best}(2014)}]{2014ARA&A..52..589H}
{Heckman}, T.~M. \& {Best}, P.~N. 2014, \araa, 52, 589

\bibitem[{{Kauffmann} {et~al.}(2003){Kauffmann}, {Heckman}, {Tremonti},
  {Brinchmann}, {Charlot}, {White}, {Ridgway}, {Brinkmann}, {Fukugita}, {Hall},
  {Ivezi{\'c}}, {Richards}, \& {Schneider}}]{2003MNRAS.346.1055K}
{Kauffmann}, G., {Heckman}, T.~M., {Tremonti}, C., {et~al.} 2003, \mnras, 346,
  1055

\bibitem[{{Kewley} {et~al.}(2001){Kewley}, {Heisler}, {Dopita}, \&
  {Lumsden}}]{2001ApJS..132...37K}
{Kewley}, L.~J., {Heisler}, C.~A., {Dopita}, M.~A., \& {Lumsden}, S. 2001,
  \apjs, 132, 37

\bibitem[{{McNamara} \& {Nulsen}(2012)}]{2012NJPh...14e5023M}
{McNamara}, B.~R. \& {Nulsen}, P.~E.~J. 2012, New Journal of Physics, 14,
  055023

\bibitem[{{Mellema} {et~al.}(2002){Mellema}, {Kurk}, \&
  {R{\"o}ttgering}}]{2002A&A...395L..13M}
{Mellema}, G., {Kurk}, J.~D., \& {R{\"o}ttgering}, H.~J.~A. 2002, \aap, 395,
  L13

\bibitem[{{Momcheva} {et~al.}(2013){Momcheva}, {Lee}, {Ly}, {Salim}, {Dale},
  {Ouchi}, {Finn}, \& {Ono}}]{2013AJ....145...47M}
{Momcheva}, I.~G., {Lee}, J.~C., {Ly}, C., {et~al.} 2013, \aj, 145, 47

\bibitem[{{Morganti} {et~al.}(2013){Morganti}, {Fogasy}, {Paragi}, {Oosterloo},
  \& {Orienti}}]{2013Sci...341.1082M}
{Morganti}, R., {Fogasy}, J., {Paragi}, Z., {Oosterloo}, T., \& {Orienti}, M.
  2013, Science, 341, 1082

\bibitem[{{Morganti} {et~al.}(1991){Morganti}, {Robinson}, {Fosbury}, {di
  Serego Alighieri}, {Tadhunter}, \& {Malin}}]{1991MNRAS.249...91M}
{Morganti}, R., {Robinson}, A., {Fosbury}, R.~A.~E., {et~al.} 1991, \mnras,
  249, 91

\bibitem[{{Mould} {et~al.}(2000){Mould}, {Ridgewell}, {Gallagher}, {Bessell},
  {Keller}, {Calzetti}, {Clarke}, {Trauger}, {Grillmair}, {Ballester},
  {Burrows}, {Krist}, {Crisp}, {Evans}, {Griffiths}, {Hester}, {Hoessel},
  {Holtzman}, {Scowen}, {Stapelfeldt}, {Sahai}, {Watson}, \&
  {Meadows}}]{2000ApJ...536..266M}
{Mould}, J.~R., {Ridgewell}, A., {Gallagher}, III, J.~S., {et~al.} 2000, \apj,
  536, 266

\bibitem[{{Neff} {et~al.}(2015){Neff}, {Eilek}, \&
  {Owen}}]{2015ApJ...802...88N}
{Neff}, S.~G., {Eilek}, J.~A., \& {Owen}, F.~N. 2015, \apj, 802, 88

\bibitem[{{Oosterloo} \& {Morganti}(2005)}]{2005A&A...429..469O}
{Oosterloo}, T.~A. \& {Morganti}, R. 2005, \aap, 429, 469

\bibitem[{{Osterbrock} \& {Ferland}(2006)}]{2006agna.book.....O}
{Osterbrock}, D.~E. \& {Ferland}, G.~J. 2006, {Astrophysics of gaseous nebulae
  and active galactic nuclei}

\bibitem[{{Rejkuba} {et~al.}(2002){Rejkuba}, {Minniti}, {Courbin}, \&
  {Silva}}]{2002ApJ...564..688R}
{Rejkuba}, M., {Minniti}, D., {Courbin}, F., \& {Silva}, D.~R. 2002, \apj, 564,
  688

\bibitem[{{Salom{\'e}} {et~al.}(2015){Salom{\'e}}, {Salom{\'e}}, \&
  {Combes}}]{2015A&A...574A..34S}
{Salom{\'e}}, Q., {Salom{\'e}}, P., \& {Combes}, F. 2015, \aap, 574, A34

\bibitem[{{Santoro} {et~al.}(2015{\natexlab{a}}){Santoro}, {Oonk}, {Morganti},
  \& {Oosterloo}}]{2015A&A...574A..89S}
{Santoro}, F., {Oonk}, J.~B.~R., {Morganti}, R., \& {Oosterloo}, T.
  2015{\natexlab{a}}, \aap, 574, A89

\bibitem[{{Santoro} {et~al.}(2015{\natexlab{b}}){Santoro}, {Oonk}, {Morganti},
  {Oosterloo}, \& {Tremblay}}]{2015A&A...575L...4S}
{Santoro}, F., {Oonk}, J.~B.~R., {Morganti}, R., {Oosterloo}, T.~A., \&
  {Tremblay}, G. 2015{\natexlab{b}}, \aap, 575, L4

\bibitem[{{Sutherland} {et~al.}(2013){Sutherland}, {Dopita}, {Binette}, \&
  {Groves}}]{2013ascl.soft06008S}
{Sutherland}, R., {Dopita}, M., {Binette}, L., \& {Groves}, B. 2013, {MAPPINGS
  III: Modelling And Prediction in PhotoIonized Nebulae and Gasdynamical
  Shocks}, Astrophysics Source Code Library

\bibitem[{{Vacca} {et~al.}(1996){Vacca}, {Garmany}, \&
  {Shull}}]{1996ApJ...460..914V}
{Vacca}, W.~D., {Garmany}, C.~D., \& {Shull}, J.~M. 1996, \apj, 460, 914

\bibitem[{{van Breugel} \& {Dey}(1993)}]{1993ApJ...414..563V}
{van Breugel}, W.~J.~M. \& {Dey}, A. 1993, \apj, 414, 563

\bibitem[{{Wagner} \& {Bicknell}(2011)}]{2011ApJ...728...29W}
{Wagner}, A.~Y. \& {Bicknell}, G.~V. 2011, \apj, 728, 29

\bibitem[{{Wagner} {et~al.}(2015){Wagner}, {Bicknell}, {Umemura}, {Sutherland},
  \& {Silk}}]{2015arXiv151003594W}
{Wagner}, A.~Y., {Bicknell}, G.~V., {Umemura}, M., {Sutherland}, R.~S., \&
  {Silk}, J. 2015, ArXiv e-prints

\bibitem[{{Weilbacher} {et~al.}(2014){Weilbacher}, {Streicher}, {Urrutia},
  {P{\'e}contal-Rousset}, {Jarno}, \& {Bacon}}]{2014ASPC..485..451W}
{Weilbacher}, P.~M., {Streicher}, O., {Urrutia}, T., {et~al.} 2014, in
  Astronomical Data Analysis Software and Systems XXIII, ed. N.~{Manset} \&
  P.~{Forshay}, Vol. 485, 451

\end{thebibliography}

\end{document}